# Precision phase measurement in Mach-Zehnder interferometer with three-photon by using a weak coherent and a squeezed vacuum state

Lifeng Duan

Department of Engineering Science, The University of Electro-Communications, 1-5-1 Chofugaoka, Chofu-shi, Tokyo, 182-8585, Japan

Email: chanter3@163.com

We quantitatively investigate phase measurement in a Mach-Zehnder interferometer (MZI), which is injected with a weak coherent and a squeezed vacuum generated from a spontaneous parametric down-conversion. The measured three-photon coincidence in this system is quantified as a function of a ratio between the amplitude of the coherent state and the squeezed parameter of squeezed vacuum when the photon number is detected at one output of MZI. It shows that the phase sensitivity reaches the Heisenberg limit when an optimal ratio is chosen. This may introduce one approach to quantify phase sensitivity in precision phase measurement.

## 1. Introduction

Precision measurement, as one of the primary studies in science, has also been developed to combat limits in modern techniques [1,2,3,4], such as sensitive instruments including microscopes[5], lithograph[6], and telescopes[7]. The measurement sensitivity in physics is fundamentally restricted by the quantum nature of Heisenberg's uncertainty principle, leading to the standard quantum limit (SQL). Since the eighties of last century, many achievements have been realized in the breaking of SQL using a non-classical state [8,9]. Finally, the sensitivity reaches Heisenberg limit by applying a set of correlated quantum entanglement [10,11,12] as probes. However, it remains a challenge to improve the absolute sensitivity since a large number quantum entanglement is difficult to generate in practice. In the case of phase measurement with Mach-Zehnder interferometer (MZI), for example, improving sensitivity below SQL of $\Delta\varphi = 1/\sqrt{\langle N \rangle}$, where $\langle N \rangle$ is the average photon for the probe light, has been achieved in numerous experiments when a squeezed vacuum inputs at the dark port. More recently, the phase sensitivity reached Heisenberg limit of $\Delta\varphi = 1/\langle N \rangle$ with an entangled NOON state as a probe [11,12].

In order to further improve absolute sensitivity in phase measurement, the generation of NOON state with a high photon number is a crucial technique. In the past decade, a number of techniques for generating different NOON states [13,14,15,16,17,18] have emerged. In general, there are two methods to produce such a state. To generate a NOON state with N photon number, one is to start from a bunch of single photon states followed by linear N-port interference and then to project measurement at the output port [13,15,19,20]. Recently experiment reported that the largest NOON state with twelve-photon was realized by combined six SPDC photon-pairs sources [18]. This technique is attractive from the view principle point, however, experimental implementation become more and more complex as N increases and lacks of successful rate because they are not simply reconfigurable. Another method is to mix a squeezed vacuum and classical coherent by choosing the optimal ratio between the amplitude of the coherent state and the squeezing parameter of squeezed vacuum [21,22,23]. A NOON states with five photon number has been experimentally demonstrated. The advantage of such a system is its simple configuration. But the experimental results show that the visibility significantly decreases as the number of photons increases. Therefore, it is necessary to further investigate such a system in quantity and to improve its capability of producing a high-photon number NOON state. On the other hand, this system has also been extensively employed to improve phase sensitivity measurement both in experiment and theory [24,25,26,27]. Since 1981, Caves [8] first proposed to beat the SQL in phase measurement, numerous schemes, such as the detection of quadrature amplitude with homodyne and detection of intensity difference between two outputs of

MZI, have been experimentally demonstrated [28]. Later, it was shown that the Heisenberg limit can also be reached when the relative number of photons at two output ports is measured. Now, it attracts more attention as one of the potential candidates for the detection of gravitational waves. For an arbitrary value of the squeezing parameter r, phase sensitivity reaches the Heisenberg scaling at $|\alpha|^2 = \sinh r$, where $\alpha$ is the amplitude of coherent state. Hence, the quantified investigation of such a system can be used to optimize parameters in future practical precision measurements. Furthermore, it also helps us to understand the principle of improving sensitivity in physics.

In this paper, we theoretically and experimentally carry out an investigation on the precision phase measurement with three-photon measurement in MZI, which is injected with a weak coherent, and squeezed vacuum state. The 3-photon rate in this system is quantified as a function of a ratio between the amplitude of the coherent state and the squeezing parameter of squeezed vacuum. It shows that the phase sensitivity reaches the Heisenberg limit when an optimal ratio is chosen. As the amplitude ratio continues to increase, the phase sensitivity gradually approaches SNL. This method provides a variable parameter—coherent amplitude for characterizing phase sensitivity in precision phase measurement, and it may be extended to quantitatively describe the precision phase with other arbitrary photon numbers.

## 2. Theoretical Model

To understand the principle, let us consider a scheme set up for investigating phase sensitivity with three-photon illustrated in Fig. 1. This setup contains three parts: input states, Mach Zehnder interferometer, and detected system. The input states, coherent state ($|\alpha\rangle$) enters into the input port $a$, and the vacuum state ($|0\rangle$) or squeezed vacuum state ($|s\rangle$) enters into another port $b$ of the Mach Zehnder interferometer. Here, our purpose is to quantify the phase sensitivity in phase measurement with three-photon, so we consider only up to three photons and neglect the contribution from more than three photons. The squeezed vacuum state and coherent state can be expanded in number state representation by $|s\rangle \approx |0\rangle - \frac{s}{\sqrt{2}}|2\rangle$ and $|\alpha\rangle \approx |0\rangle + \alpha|1\rangle + \frac{\alpha^2}{\sqrt{2}}|2\rangle + \frac{\alpha^3}{\sqrt{6}}|3\rangle$. The coherent state has a complex amplitude of $\alpha = |\alpha|e^{i\theta}$ and the squeezed vacuum state has a squeezing parameter of $s$, which is real small positive number. After BS1, the beams $c$ and $d$ are reflected by a pair of mirrors onto the second beam splitter (BS2). We only concern on the case of three photon, so the output state of Mach-Zehnder interferometer can be written by

$$|\psi_{out}\rangle = \frac{1}{4\sqrt{3}}\left((\alpha^2 - 3s)\alpha(1 + e^{3i\phi}) + 3(\alpha^2 + s)\alpha(e^{2i\phi} + e^{i\phi})\right)|3,0\rangle$$

$$+ \frac{1}{4\sqrt{3}}\left((\alpha^2 - 3s)\alpha(1 - e^{3i\phi}) - 3(\alpha^2 + s)\alpha(e^{i\varphi} - e^{2i\varphi})\right)|0,3\rangle$$

$$+ \frac{1}{4}\left((\alpha^2 - 3s)\alpha(1 - e^{3i\phi}) + (\alpha^2 + s)\alpha(e^{i\phi} - e^{2i\phi})\right)|2,1\rangle$$

$$+ \frac{1}{4}\left((\alpha^2 - 3s)\alpha(1 + e^{3i\phi}) - (\alpha^2 + s)\alpha(e^{i\varphi} + e^{2i\varphi})\right)|1,2\rangle$$

(1)

where $\phi$ is the phase shift between beam $c$ and $d$. We use the method of projective measurement to obtain three-photon probability of one output port. After Mach-Zehnder interferometer, the three-photon rate at output $f$ can be expressed like this,

$$R_{30} \propto A + B\cos3\phi + C\cos2\phi + D\cos\phi, \qquad (2)$$

in which

$$A = |\alpha(\alpha^2 - 3s)|^2 + 9|\alpha(\alpha^2 + s)|^2,$$
$$B = |\alpha|^2(\alpha^2 - 3s)^2,$$
$$C = 6|\alpha|^2(\alpha^2 - 3s)(\alpha^2 + s),$$

$$D = 3|\alpha|^2(\alpha^2 + s)(5\alpha^2 - 3s). \tag{3}$$

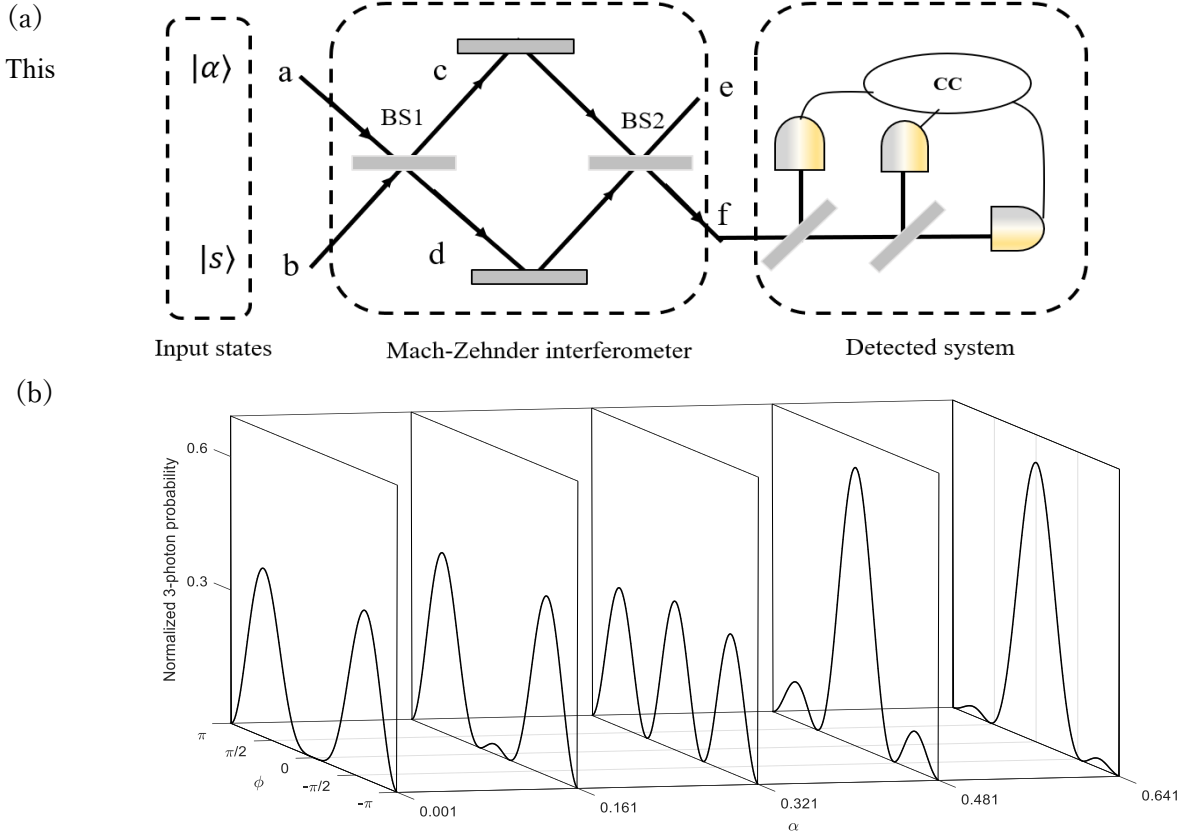

FIG.1. (a) Schematic setup diagram for investigating phase sensitivity with three-photon (b) Theoretical simulation diagram of the normalized 3-photon probability as a function of coherent amplitude $\alpha$ and MZ phase $\phi$, when we set $s$=0.1.

equation indicates that the three-photon coincidence rate is a function of phase shift ($\varphi$) and the amplitude of coherent and squeezed vacuum state. If the squeezing parameter $s = 0$, it indicates the input state is coherent plus vacuum state, and the three-photon coincidence rate has the form of $R_{30} \propto C(1 - cos\varphi)^3$, whose phase sensitivity is limited by classical SNL. For a given pump power with squeezing parameter $s$, if we set $\alpha^2 = -s$, the coefficients in front of $cos2\varphi$ and $cos\varphi$ are zero and the three-photon rate of $R'_{30} \propto 1 + cos3\varphi$ will be obtained. This can be understood that three-photon N00N state can be obtain in Mach Zehnder interferometer by choosing the coherent intensity. In this condition, the phase sensitivity can reach Heisenberg limit.

To further understand the obtained three-photon rate, we normalized the three-photon rate by the means of $\int_{-\pi}^{\pi} P_{30} d\phi = 1$. The theoretical simulation diagram of normalized 3-photon probability is shown in FIG.1(b). For a given pump power ($s = 0.1$) and the relative phase between coherent and squeezed vacuum state is set to be PI, the normalized three-photon probability varied by MZ phase with different coherent intensity. As can be seen, when the coherent intensity is tiny ($|\alpha|^2 \ll s$), the three-photon probability has two periods. It indicated that there has no three-photon probability generate in coherent and the three-photon probability is from a new three-photon state, which composed of the two-photon from squeezed vacuum state and one-photon from coherent. Therefore, the oscillation period is mainly characterized by squeezed vacuum state and the phase sensitivity is sub-SNL. As the intensity of the coherent light increases ($|\alpha|^2 < s$), a small oscillation period appears in the middle of the two oscillation periods and the middle oscillation period increases with coherent intensity increase. It indicates that, when the three-photon state exists in coherent, the three-photon state of the coherent interference with the new three-photon state on the first beam splitter of the interferometer, forming a three-photon-like NOON state, which contains not only three-photon NOON state components, but also non-NOON state components. The phase sensitivity continues to decrease because of the increasing gradually of three-photon NOON state composition in Mach-Zehnder interferometer. As the coherent light increases, especially when $\alpha^2 = s$, there has three standard oscillation periods, indicating that there only exist three-photon N00N state in Mach-Zehnder interferometer and the phase sensitivity is considered to reach Heisenberg limit. For $\alpha^2 > s$, the middle period gradually larger than the other two periods. the phase sensitivity gradually increases, there has non-pure three-photon N00N state in Mach-Zehnder interferometer can be explained, so the phase sensitivity gradually away from Heisenberg limit. With the sustained increasing of the coherent intensity, the phase sensitivity gradually approaches the standard quantum limit. The result is consistent with phase sensitivity when there has only coherent light [20].

## 3. Experimental setup

To investigate phase sensitivity with three-photon, a layout of the experimental setup is shown in Fig. 2. Our experimental setup mainly contains three parts. The first part is the generation of squeezed vacuum state, which is shown in Fig.2(a). It is similar to our previous experiment for measuring three-photon interference between a two-photon state and a weak coherent state on a beam splitter [28]. The second part is Mach-Zehnder interferometer, which is shown in Fig.2(b), contains beam splitter1(BS1) and beam splitter 2(BS2). The third part is the detected system, which is shown in Fig.3(c), consists of couplers, single-mode fiber beam splitters (Thorlabs FC830-50B-FC), single photon counting modules (SPCM, EXCELITAS, SPCM-AQRH13-FC), an electronical coincidence circuit and photon counters.

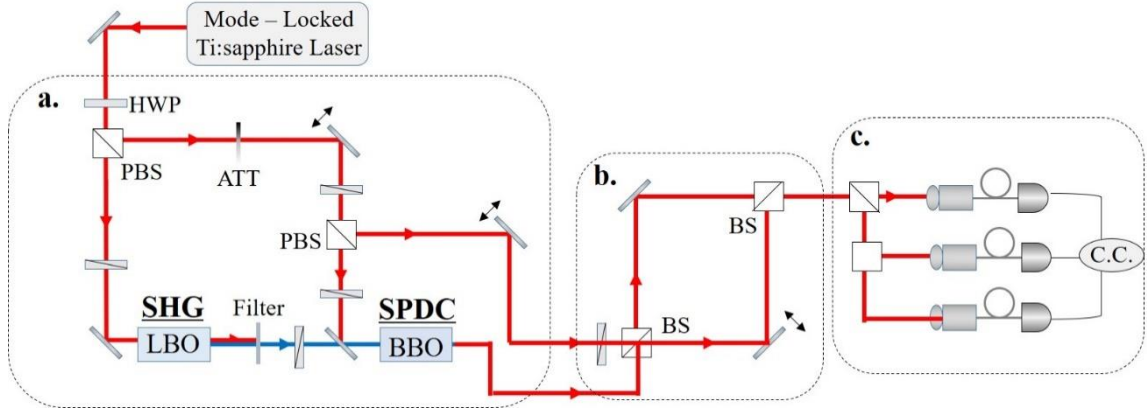

FIG.2. Experimental setup for investigating phase sensitivity. (a) Generation layout of squeezed vacuum state and preparation of the input states(coherent and squeezed vacuum state); (b) Mach-Zehnder interferometer; (c) Detected system.

SHG: Second harmonic generator; SPDC: Spontaneous parametric down conversion; HWP: Half wave plate; PBS: polarization beam splitter; SPF: Short wave-pass filter

A continuous-wave mode-locked Ti: Sapphire laser with 2 ps pulse duration at a pulse repetition rate of 82 Mhz, was employed as primary source. Most of the 798 nm fundamental light ($H$ polarization) is frequency doubled by using a single-pass-through 15 mm long Type-I Lithium triborate (LBO) nonlinear crystal to obtain 399 nm ultraviolet light ($V$ polarization) with the maximum power of 50 mW. The 399 nm pump pulses are mixed with the remaining fundamental light ($V$ polarization) and injected into the 5 mm*5 mm*5 mm beta barium borate (BBO) crystal. The 399 nm light is employed to pump collinear degenerated Type-I spontaneous parametric down conversion (SPDC) at 798 nm using the BBO crystal. A squeezed vacuum state ($H$ polarization), with a squeezing parameter of 0.06(minimum de-amplification 0.89, maximum amplification 1.13) at the pump power of 40mW, was generated, when the SPDC is adjusted to be optical parametric amplifier simultaneously.

To ensure the remaining fundamental light reached the PDC simultaneously with the 399nm pump light, an optical temporal delay was introduced by a movable prism, which can be controlled by stepping motor. Meanwhile, the relative phase between the two beams was also finely adjusted and stabilized by a piezoelectric translator1(PZT1), which was mounted on a mirror. we adjust HWP1 to make the polarization of coherent, which worked as a reference light, is orthogonal with pump. The squeezing parameter can be estimated by operating the system as an optical parametric amplifier when the coherent worked as a reference. The two same polarization squeezed vacuum beam and weak coherent, acted as the input states, are injected into the Mach-Zehnder interferometer, respectively.

In second part of the Mach-Zehnder interferometer, consisted of BS1 and BS2. The three-photon interference of weak coherent state and squeezed vacuum state have occurred on BS1. This means make there has no vacuum state occurs in the Mach-Zehnder interferometer. The relative phase between the two paths in the interferometer was adjusted by a mirror, which was also mounted on a PZT2. The investigation of phase sensitivity in this experiment is partly dependent on the stability of the relative phase between the coherent light and the pump. In order to improve the stability, all optical components for SHG and coherent light were arranged on one breadboard.

After the Mach-Zehnder interferometer, the three-photon counting rate of one output port of the Mach-Zehnder interferometer was investigated with our detection systems. The observed single count rates of the produced squeezed vacuum state were 30 kcps and its two-fold coincidence counting rates was about 2.8 kcps when the squeezing parameter is 0.06. Compared with single counting rates, the dark count of 200 cps for SPCM is so small that it can be neglected. The intensity of coherent was varied by changing attenuation, which is realized by a variable transmittance filter in the coherent beam.

## 4. Experimental results and discussion

Fig.4 shows the observed three-fold coincidence counting rate as a function of the MZ phase in the Mach-Zehnder interferometer at the amplitude ratio $\alpha^2/s = 0.3, 0.53, 0.89, 1.03, 1.19, 4.7$, between the weak coherent state and the squeezed vacuum state, when the squeezed vacuum state squeezing parameter is $s = 0.06$. The measured dates are fitting with the equation of $R' = A\big(B - (y-3)^2 \cos 3(\varphi - \varphi_1) + 12(y-3)(y+1)\cos 2(\varphi - \varphi_1) - 6y(y+1)(5y-3)\cos 2(\varphi - \varphi_1)\big)$, which is from the analytical model in Eq. (2). Here, $y = a\,\alpha^2/s$, where $a$ is used to scale the measured date and $\varphi_1$, which is the phase shift caused by the unstable of experiment system. In experiment, it was realized by two steps. In the first step, the three-photon interference is occurred in BS1, which composed of HWP2 and PBS2, between the three-photon probability amplitude originated from coherent and combined three-photon probability from the two-photon state and the a single-photon in the coherent and the three-photon NOON state is generated in MZ interferometer. The composition of the three-photon NOON state in MZ interferometer changes with coherent intensity due to the three-photon interference. In the second step, phase shift (MZ phase) exists etween the two arms in MZ interferometer. After BS2, the phase sensitivity was investigated by measuring the three-photon coincidence rate and the MZ phase was scanned. It can be seen from Fig.4. that the oscillation pattern has different visibility and the maximum visibility is about $(48 \pm 5)\%$. These imperfect visibilities may be caused by spectrum difference, because the spectrum of the two-photon from SPDC is much wider than the coherent. Meanwhile, the spatial mis-alignment and temporal mismatching between the squeezed vacuum state and classical coherent also make the visibility decreased. With an increment of the intensity of coherent, the oscillation period of three-photon counting rate gradually in the range of 0-2π changes from two-times to three-times and finally to one time. For the case of $\alpha^2/s = 1.03$, there has three standard period in the range of 0-2PI. When the coherent amplitude is large enough, it is difficult to observe because of the three-photon counting rate is dominated by a three-photon state in coherent.

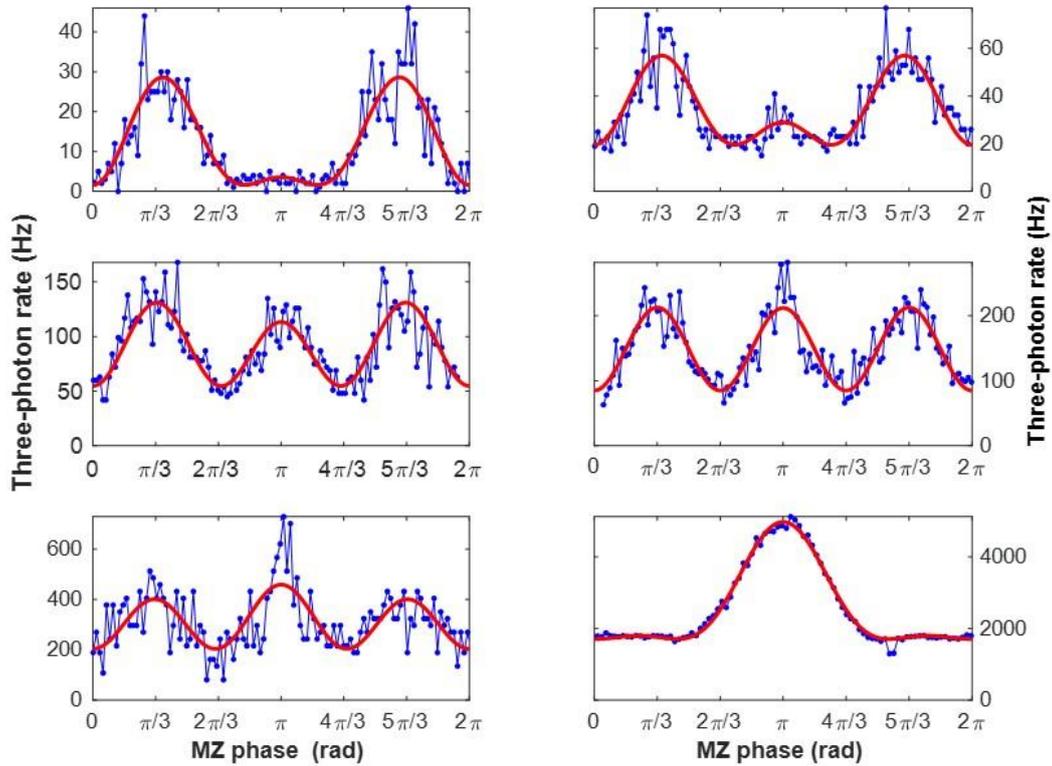

Fig.3. Experiment results of three-photon rate as a function of MZ phase at different amplitude of $\alpha^2/s = \alpha^2/s = 0.3,\ 0.53,\ 0.89,\ 1.03,\ 1.19,\ 4.7$

To further understand phase sensitivity with three-photon, we can consider to analyze the above discussed behaviors like this. For $\alpha^2/s < 1$, the phase sensitivity monotonically decreases with the increment of the amplitude ratio between input. This indicates that the phase sensitivity gradually approaches to the Heisenberg limit from sub-standard quantum limit(sub-SQL), because of the existence of squeezed vacuum state. For $\alpha^2/s = 1$, the three-photon NOON state is generated and the phase sensitivity can reach the Heisenberg limit. When $\alpha^2/s > 1$, the phase sensitivity gradually increases, indicating that there has other non-NOON state component and the phase sensitivity moves gradually away from the Heisenberg limit and close to SQL. When $\alpha^2/s \gg 1$, the phase sensitivity infinitely closes to SNL and shows the same decreasing trend with SQL.

## 5.Conclusion

In conclusion, we theoretically proposed and experimentally investigated phase sensitivity with three-photon in Mach-Zehnder interferometer by injecting a weak coherent and squeezed vacuum state generated from a spontaneous parametric down conversion. Most of previous reports pay more attention to make precision phase measurement by choosing optimal input amplitude ratio to generate ideal high-photon NOON state and make the phase sensitivity reach the fundamental Heisenberg limit. This study from another perspective, the phase sensitivity is quantified as a function of the optical ratio of between coherent amplitude and the squeezing parameter of squeezed vacuum state. The result shows that the phase sensitivity reaches the Heisenberg limit when an optimal amplitude ratio is chosen. Furthermore, this quantified means may find a fundamental connection between SQL and Hl. This is the first time to quantitatively describe phase sensitivity with photon-number detection by injecting quantum state and classical coherent state. In addition, for further extending the measurement of high photon phase sensitivity, we only consider the correlation of more photons and the choice of weak coherent light intensity, rather than preparation of high squeezing light field. To summary, we believe this approach may be useful for measuring precision phase in quantum metrology, quantum imaging and sub-Rayleigh lithography.